\documentclass{article}
\usepackage{setspace}
\usepackage{arxiv}
\usepackage{algorithm}
\usepackage{algpseudocode}
\usepackage[utf8]{inputenc} 
\usepackage[T1]{fontenc}    
\usepackage{hyperref}       
\usepackage{url}            
\usepackage{booktabs}      
\usepackage{amsfonts}       
\usepackage{nicefrac}       
\usepackage{microtype}      
\usepackage{lipsum}
\usepackage{graphicx}
\usepackage{amsmath}
\usepackage[dvipsnames]{xcolor}
\definecolor{darkblue}{RGB}{0,0,139}
\definecolor{midnightblue}{RGB}{25,25,112}
\usepackage{caption}
\usepackage{bm}
\usepackage[font={color=darkblue},figurename=Figure]{caption}
\graphicspath{ {./figures/} }
\usepackage{authblk}
\usepackage[version=4]{mhchem}

\title{Controlled Ion Transport in the Subsurface: A Coupled Advection-Diffusion-Electromigration System}
\author[a,*]{Kunning~Tang}
\author[b]{Zhenkai~Bo}
\author[c]{Zhe~Li}
\author[a]{Ying Da~Wang}
\author[d]{James McClure}
\author[e]{Hongli~Su}
\author[a]{Peyman~Mostaghimi}
\author[a,†]{Ryan T.~Armstrong \thanks{Corresponding author: Ryan T. Armstrong, Email: ryan.armstrong@unsw.edu.au}}

\affil[a]{School of Minerals and Energy Resources Engineering, The University of New South Wales, Sydney, NSW 2052, Australia}
\affil[b]{School of Chemical Engineering, The University of Queensland, Brisbane, QLD 4072, Australia}
\affil[c]{Research School of Physics, The Australian National University, Canberra, ACT 2601, Australia}
\affil[d]{National Security Institute, Virginia Tech, Blacksburg, VA 24061, USA}
\affil[e]{Institute of Frontier Materials, Deakin University, Geelong, VIC 3220, Australia}

\begin{document}
\setcounter{tocdepth}{4}
\setcounter{secnumdepth}{4}
\maketitle

\pagebreak
\begin{abstract}

Groundwater pollution poses a significant threat to environmental sustainability during urbanization. Existing remediation methods like pump-and-treat and electrokinetics have limited ion transport control. This study introduces a coupled advection-diffusion-electromigration system for controlled ion transport in the subsurface. Using the Lattice-Boltzmann-Poisson method, we simulate ion transport in various two- and three-dimensional porous media. We establish an ion transport regime classification based on the P\'eclet number ($Pe$) and a novel Electrodiffusivity index ($EDI$). By manipulating the electric potential, hydrostatic pressure, and ion concentration, we identify four transport regimes: large channeling, uniform flow, small channeling, and no flow. Large channeling occurs when advection dominates, while uniform flow arises when diffusion and electromigration are more prevalent. Small channeling happens when the advection opposes electromigration and diffusion, and no flow occurs when the advection or electromigration impedes ion transport via diffusion. Simulations in heterogeneous models confirm these transport regimes, highlighting the influence of pore size variation on transport regimes. Consequently, $Pe$ and $EDI$ must be tailored for optimal transport control. These findings enable better control over ion transport, optimizing processes such as heavy metal removal, bioremediation, and contaminant degradation in groundwater management.
\end{abstract}

\textbf{Key Points}

1. Advection-Diffusion-Electromigration is coupled to understand the controlled ion transport in porous media using the scaling analysis.

2. A transport diagram with four regimes is obtained based on the P\'eclet number and a novel Electrodiffusivity index in 2D and 3D porous media.

3. The transport diagram provides guidance to control the subsurface ion transport by adjusting the P\'eclet number and Electrodiffusivity index.


\pagebreak
\section{Introduction}
\label{sec:intro}


Sediments and aquifers are fundamental to global environmental sustainability and provide most of the freshwater reserves on earth ($98-99\%$) \cite{lima2017environmental,fensham_response_2022}. As such, groundwater management, which deals with complex interactions between human activities and the corresponding subsurface environment, is crucial for the development of our community \cite{sen_chapter_2015,fensham_response_2022,inglis_electrokinetically-enhanced_2021,wang_agent-assisted_2022}. During modern urbanization, pollution is widely recognized as a significant challenge to soil and groundwater \cite{nations_groundwater_2003,mackay_groundwater_1989,lima2017environmental,sanganyado_organic_2021}. Common pollutants range from soluble chemicals to organic compounds \cite{virkutyte_electrokinetic_2002,sanganyado_organic_2021,wang_agent-assisted_2022,chen_emerging_2022}, thus corresponding remediation requires the removal of hazardous ions and organic molecules from soil and water. Conventional groundwater remediation methods such as pump-and-treat started in the last century aim to direct the ion flux to a designated direction via hydrostatic advection of fluids\cite{mackay_groundwater_1989}. Moreover, other groundwater management technology, such as dewatering, also involves manipulating ion flows \cite{patel_4_2019,li_enhancing_2023,ji_fouling_2023}. When implementing these technologies, ions dissolved in water or absorbed on the soil surface are subject to advection and diffusion flow mechanisms \cite{sanganyado_organic_2021,bedrikovetsky_analytical_2014}. However, soil and sediments are naturally heterogeneous, leading to unevenly distributed pressure and ion concentration gradients along flow paths \cite{ringrose_reservoir_2015,ding_forward_2022}. In such circumstances, ions will predominantly move along the flow path with the lowest pressure and concentration gradient, named channeling flow \cite{li1995scaling,farajzadeh_effect_2011}. Therefore, it is difficult to control the ion fluxes in the subsurface solely by harnessing advection and diffusion flow mechanisms \cite{mackay_groundwater_1989,virkutyte_electrokinetic_2002,martens_toward_2021,ding_forward_2022}. To resolve this issue, electrokinetic (EK) transport is proposed, which applies an electric potential to initiating ion fluxes toward a designated direction \cite{virkutyte_electrokinetic_2002,casagrande1949electro,inglis_electrokinetically-enhanced_2021}. A unique benefit of EK is enhancing the accessibility of ions into low-permeable regions where long-term resources or pollutants are often trapped \cite{sprocati2022interplay}. However, the energy expenditure for removing pollutants with EK is usually high because the EK process is slow, and successful remediation can take many years \cite{van1997electrokinetics,virkutyte_electrokinetic_2002}. Therefore, we propose that a coupled advection-diffusion-electromigration flow system can better control the ion fluxes to achieve an optimal flow behavior, and relevant work is needed to prove this assumption.

\begin{figure}[htp!]
  \centering
    \includegraphics[width=1\textwidth]{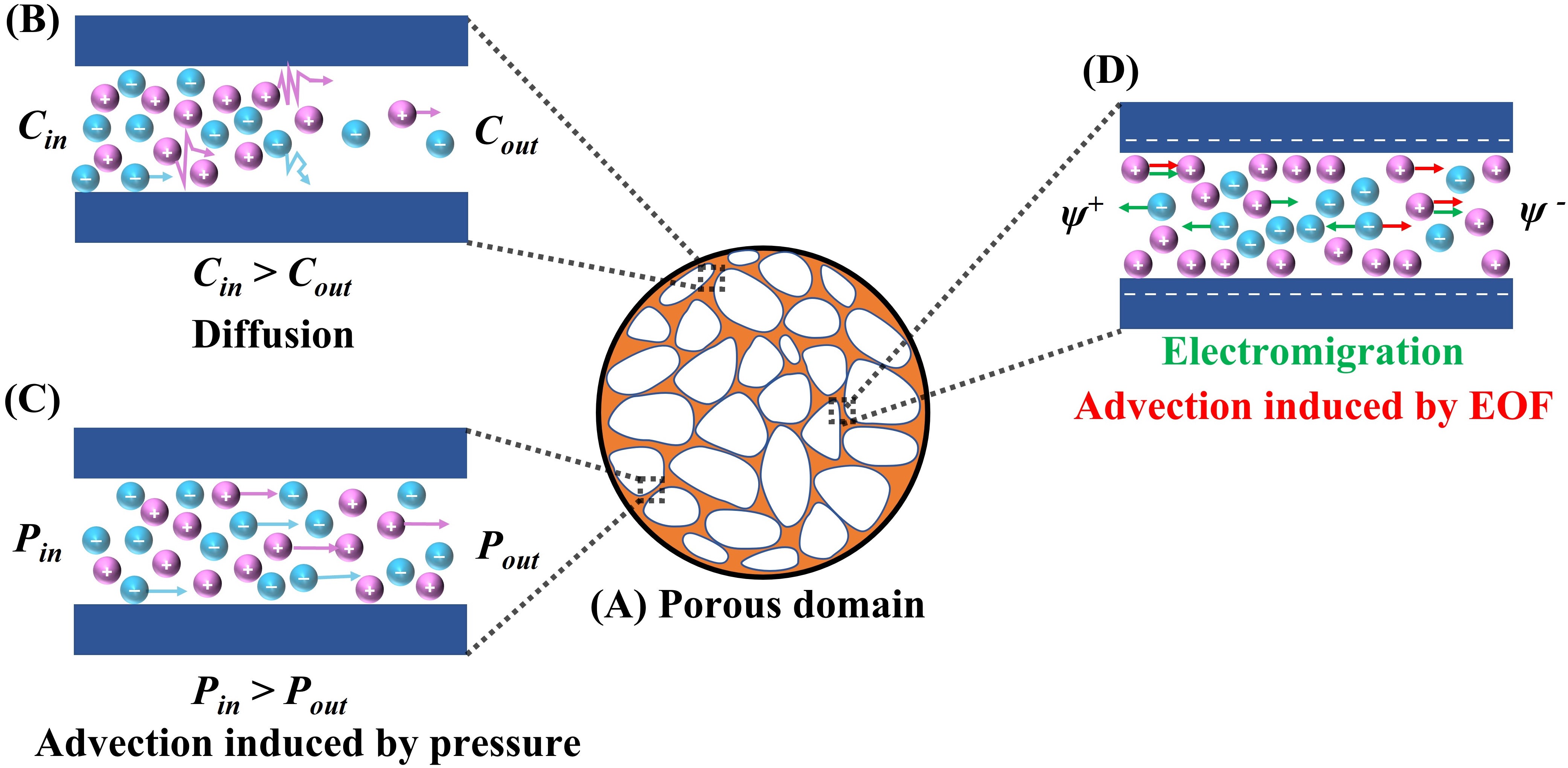}
    \caption{Schematic description of each ion driving mechanism. (A) A porous domain; (B) Ion transport due to diffusion; (C) Ion transport due to advection which is induced by external hydraulic pressure; (D) Ion transport by electromigration due to an external electric field (marked by green arrows), and advection due to EOF with a charged solid surface (marked by red arrows).}
    \label{fig:Mechanisms}
\end{figure}

Under the advection-diffusion-electromigration system, ion flow is affected by several important factors: diffusion, pressure-induced advection, advection due to electroosmosis, and electromigration \cite{martens_toward_2021,sprocati2022interplay}. These mechanisms are described in Figure \ref{fig:Mechanisms}. Ions (molecules/particles) in the fluids are initially subject to gravity force; meanwhile, the fluid motion will exert drag force due to hydrostatic pressure (advection of fluids) \cite{bedrikovetsky_analytical_2014}. Because of the uneven concentration distribution, ions are also driven by diffusion, where the diffusion rate depends on ion type and porous media tortuosity \cite{shen_critical_2007}. EK transport happens after an electric potential is applied, which consists of electromigration and electroosmosis \cite{yuan_effect_2023}. While electromigration involves the displacement of charged particles/ions towards an electrode of opposite charge, the electrical double layer (EDL) plays a key role in electroosmotic flow (EOF) \cite{zhang_electro-osmosis_2017}. A charged solid surface and a thin layer of counter ions in an aqueous solution form an EDL. As counter ions in the EDL move towards the oppositely charged electrode, momentum is transferred to the surrounding fluid molecules, inducing ion flux \cite{acar1993principles}. The Helmholtz--Smoluchowsky equation (HS) is commonly used to determine EOF in porous media, and its application depends on the thickness of the EDL \cite{acar1993principles, WANG2007264,zhang_electro-osmosis_2017}. Most studies in the literature assume a thin double layer, which means that the thickness of the EDL is considerably smaller than the pore size \cite{WANG2007264, zhang_electro-osmosis_2017}.  In such cases, the zeta potential is a way to characterize the EDL based on the ionic concentration in the EDL and the pH as a result of the protonation/deprotonation reactions at the particle surface \cite{VANE19971, Lima,zhang_electro-osmosis_2017, KHOSO201955, LIU2010542}.

Other essential factors to be considered during electrokinetics are local pH distribution, and the pore-scale characteristics of the porous media (e.g. tortuosity and porosity) \cite{mattson2002electrokinetic,appelo2007multicomponent,al2008electrokinetic,storey2012effects,zhang_electro-osmosis_2017,sprocati2019modeling,sprocati2020charge,priya2021pore}. Local pH heterogeneity causes the heterogeneity of zeta potential and results in a nonlinear response of the electroosmotic velocity. Moreover, tortuosity and porosity provide the morphological information of the porous media and provide the bridge to study the  micro-scale and macro-scale relationship for future upscaling \cite{yeung1994chapters,pengra1995electrokinetic,pengra1999determination}. Alizadeh et al.\cite{alizadeh2019impact,sprocati2022interplay} studied the effect of the porosity heterogeneity on EK transport and found that heterogeneities play an important role in EK transport and the coupling to the hydraulic process. Therefore, to accurately model EOF and subsequent EK transport, the thickness of the EDL and zeta potential must be characterized based on realistic chemical conditions and porous media morphology. However, most previous studies on EOF and electrokinetics are based on simplified pore-structure models \cite{WANG2007264,wang_lattice_2006,zhang_electro-osmosis_2017,Alizadeth}. These studies mainly focused on the fundamental physical perspective of EOF and used manually generated porous media as the study domain and the effect of EOF on the flow behavior in complex and realistic porous media has yet to be investigated \cite{chen_emerging_2022}.

Due to the lack of related work and the technology limitation, it is difficult to accurately estimate the actual dimension of polluted regions in the subsurface \cite{lima2017environmental,zaporozec_groundwater_2004}. Thus, it will be physically impossible to predict the ion transport behavior (pressure, velocity, ion concentration distribution) at a larger scale, e.g. hundred to thousands of meters\cite{yang2017forward}. The non-dimensional scaling theory, which refers to translating results from one scale (smaller) to another (larger) under the same dimensionless groups, has been proposed and widely used to solve problems in fluid mechanics \cite{buckingham_physically_1914,shook_scaling_1992}. The ion transport behavior is commonly described by the dimensionless number P\'eclet number ($Pe$, the ratio between ion fluxes due to advection and diffusion) \cite{steefel_multicomponent_1998,simmons_modelling_1998,mostaghimi2016numerical}. \cite{steefel_multicomponent_1998} conducted a numerical simulation for reactive transport in a dual-porosity porous system validated by analytical solution, finding that the geometry of reaction font during simulation can be characterized by a revised form of the P\'eclet number. \cite{chen_generalized_2012} studied the ion fluxed behaviors in the finite and semi-infinite domains where different P\'eclet numbers (advection or dispersion dominant flow) will significantly affect the ion breakthrough concentration profiles at effluent boundaries. P\'eclet number is also used to describe the dispersion coefficient resulting from advection-diffusion coupled flow in pore-network modeling where electroosmosis is involved  \cite{li_solute_2014}. Moreover, scaling theory has been well applied in the oil and gas industry for predicting recovery and ion sweep for decades \cite{li1995scaling,shook_scaling_1992,wang_scaling_2022}. Despite the importance of scaling theory in the fluid flow in porous media and various engineering problems, to our best knowledge, there has not been a systematic scaling analysis study on the advection-diffusion-electromigration transport system, not to mention the corresponding transport behavior and implications on potential field practice \cite{martens_toward_2021}.

From the literature, one can conclude that many important factors can affect ion flows in the subsurface \cite{bedrikovetsky_analytical_2014,ding_forward_2022}. However, their coupled influences and controls on ion fluxes have never been completely investigated \cite{shen_critical_2007,li1995scaling,wang2022scaling}. Full-physics simulation of their coupling between different mechanisms is therefore needed to understand the ion fluxes. Herein, the aim is to conduct a numerical-based scaling analysis of ion transport regimes under electric potential, pressure field, and ion diffusion in subsurface porous media and the effect of pore-space complexity on the ion fluxes. We use the P\'eclet number ($Pe$) to characterize the ion fluxes resulting from diffusion and advection. Similarly, to describe the ratio between ion fluxes due to electromigration and diffusion, we define a new dimensionless number called the Electrodiffusivity index ($EDI$). Then, we benchmark the ion flux flow regimes according to visual observations from flow simulations in a 2D microchannel model under various electric potentials, pressure fields, and concentration conditions. Further numerical simulations are conducted in a 2D heterogeneous porous media and a 3D heterogeneous domain built from a micro-computed Tomography (micro-CT) sandstone image to validate the benchmarked ion flow regimes observation from the microchannel model. Lastly, we investigate the heterogeneity or tortuosity effects on the ion flow regime under the same conditions. All the simulations are carried out in our open-source solver OPM/LBPM using the lattice Boltzmann-Poisson method \cite{tang2023pore}. We benchmark four ion flux flow regimes from the simulation in a 2D microchannel model: large channeling, uniform flow, small channeling, and no flow. The same flow patterns are also observed during simulations in the 2D and 3D heterogeneous domains. Furthermore, to facilitate future applications in the field, we quantify the flow patterns in 2D and 3D heterogeneous domains by studying ion transport efficiency and time until ion breakthrough. The results show that the ion flow can be well-controlled by changing the electric potential, pressure field, and concentration difference. Overall, our results demonstrate how ion transport can be controlled in heterogeneous environments by adjusting the external electric potential, hydrostatic pressure, and injected ion concentration. Many applications, such as heavy metal removal in soil, bioremediation, and contaminant degradation for groundwater management \cite{yuan_electrokinetic_2006,virkutyte_electrokinetic_2002,inglis_electrokinetically-enhanced_2021,ding_forward_2022,chen_emerging_2022}.

\section{Methods and Materials}
\label{sec:Methods and ore sample description}

\subsection{Theoretical Background and Numerical Methods}
\label{sec:Numerical}

The governing equation for the advection-diffusion-electromigration flow includes three coupled equations: Nernst--Planck equation for ion transport driven by chemical and electric potentials, Navier--Stokes equation for the flow of an electrolyte solution carrying ions driven by hydrodynamic potential, and Poisson equation that couples ionic distribution to compute electric potential, which in turn is fed into Nernst--Planck and Navier--Stokes equations.

More specifically, the transport of an ion species is modeled by the Nernst--Planck equation:
\begin{equation}\label{eq:NernstPlanckEq}
    \frac{\partial C_i}{\partial t} + J \cdot \bm{J}_i = 0, 
\end{equation}
where $C_i$ is the concentration of the $i$th ion, and $\bm{J}_i$ is the associated mass flux, given by:
\begin{equation}\label{eq:flux_terms}
    \bm{J}_i = \bm{J}_i^d + \bm{J}_i^a + \bm{J}_i^e,
\end{equation}
where the net diffusive flux, $\bm{J}_i^d$, advection flux, $\bm{J}_i^a$, and electromigration flux, $\bm{J}_i^e$ are defined as:
\begin{equation}
    \begin{aligned}
        \bm{J}_i^d &= -D_i \nabla C_i, \\
        \bm{J}_i^a &= \bm{u} C_i, \\
        \bm{J}_i^e &= - \frac{z_i D_i}{V_T} \nabla \psi C_i,
    \end{aligned}
\end{equation}
where $D_i$ is the diffusivity of the $i$th ion, $z_i$ is the ion algebraic valency, and $V_T=k_B T/e$ is the thermal voltage, where $k_B$ is the Boltzmann constant and $e$ is the electron charge.

The Nernst--Planck equation is solved in the pore spaces, and a non-flux boundary condition was applied to model the bounce-back of ions of the ion--solid interface:

\begin{equation}\label{eq:non-flux_BC_ion}
    \bm{n}_s \cdot \bm{J}_i = 0,\;\; \text{for}\; \bm{x} \in \partial \Omega,
\end{equation}
where $\bm{n}_s$ is the unit normal vector of the solid surface. 

The flow of the electrolyte solution that carries ion species is governed by the incompressible Navier--Stokes equations and the conservation of mass:

\begin{equation}\label{eq:NavierStokesEqs}
    \begin{aligned}
        \nabla \cdot \bm{u} &= 0, \\
         \frac{\partial \bm{u}}{\partial t} + \bm{u} \cdot \nabla \bm{u} &= - \frac{1}{\rho_0}\nabla p + \nu \nabla^2 \bm{u} + \frac{\bm{F}}{\rho_0},
    \end{aligned}
\end{equation}

where $\bm{u}$ is the fluid velocity vector, $\rho_0$ is the fluid density, $p$ is the fluid pressure, $\nu$ is the kinematic viscosity, and $\bm{F}$ is the body force, which in this study is essentially due to an external electric field. 

The standard no-slip boundary condition is applied to the fluid-solid interface. In addition, when the thickness of the EDL is much smaller than the characteristic length of the simulation domain (i.e. the resolution of an input image), an electroosmotic velocity boundary condition is introduced in the EDL, which ignores the detailed flow between the solid surface and slipping plane and analytically calculates the velocity at the solid according to the local zeta potential of the solid surface. In this work, we adopted the widely used Helmholtz-Smoluchowski (HS) equation:

\begin{equation}\label{eq:HSequation}
    \bm{u} = - \frac{\epsilon \zeta}{\mu} \nabla_T \psi,\;\; \text{for}\; \bm{x} \in \partial \Omega,
\end{equation}

where $\psi$ is the electric potential within the fluid and $\Omega$ denotes the fluid domain, $\zeta$ is the local zeta potential at the solid surface, $\epsilon$ is the permittivity of the electrolyte solution, and $\nabla_T$ is the tangential part of the gradient operator, perpendicular to the orientation of the solid surface. In short, when EDL is not resolvable due to the input image resolution, the electroosmotic velocity boundary is applied to replace the electrical body force in Eq.\ref{eq:NavierStokesEqs}.


The electric potential due to the ion transport is solved by the Poisson equation:

\begin{equation}
    \nabla^2 \psi = - \frac{\rho_e}{\varepsilon_r \varepsilon_0},
\end{equation}
where $\varepsilon_0$ is the permittivity of vacuum, and $\varepsilon_r$ is the dielectric constant of the electrolyte solution. The net charge density $\rho_e$ (C/m$^3$) is calculated based on the ion concentration:

\begin{equation}
    \rho_e = \sum_i F z_i C_i,
\end{equation}
where the sum runs over all ionic species and $F$ is Faraday's constant given by $F=e N_A$, where $N_A$ is Avogadro's number. The Poisson equation couples the Nernts--Planck and Navier--Stokes equations together, as it takes charge concentration as an input, solves for the electric potential that in turn affects ion transport, and leads to a Coulomb force in the fluid transport. For regions where the net charge is non-zero, an induced electrical body force that can drive the fluid flow in the Navier--Stokes equation is given by:

\begin{equation}
    \bm{F}_e = \rho_e \bm{E} =- \rho_e \nabla \psi.
\end{equation}

Normally a non-zero charge distribution occurs in EDL; when the EDL is not resolvable, the body force $\bm{F}_e$ reduces to the HS boundary condition in Eq.\ref{eq:HSequation}. At the solid surface, the boundary condition for electric potential is typically specified in two forms: (a) the surface charge density $\sigma_e$ and (b) a surface potential $\psi_s$  (when the EDL is unresolvable, the surface potential reduces to the zeta potential). The former is a Neumann-type boundary condition given by

\begin{equation}
    \bm{n}_s \cdot \nabla \psi = - \frac{\sigma_e}{\varepsilon_r \varepsilon_0}, \;\; \text{for}\; \bm{x} \in \partial \Omega,
\end{equation}
and the latter is a Dirichlet-type boundary given by

\begin{equation}
    \psi(\bm{x}) = \psi_s,\;\; \text{for}\; \bm{x} \in \partial \Omega,
\end{equation}

where $\psi_s$ is a user-specified electric potential of the solid surface. To solve all the coupled equations above in porous media, we adopted the commonly used numerical method, the lattice-Boltzmann method (LBM), thanks to its inherent scalability of parallel computation and efficient handling of complex boundary conditions. The numerical details of the LBM model implemented in this work and the validation of governing equations can be found in \cite{mcclure2021lbpm,tang2023pore}. The open-source code for the model is accessible on GitHub (https://github.com/OPM/LBPM). Simulations were performed on a local workstation with a 64-core CPU, 24 GB of GPU memory, and 256 GB of RAM. All simulations were computed on the GPU due to a faster computational speed than the CPU.

\subsection{Simulation of Domain-Synthetic Porous Media}
\label{sec:Domain-Synthetic}

Electrokinetic technology application environments range from unconsolidated soil to consolidated sedimentary rocks where porous media heterogeneity impacts the transport behaviors. To conduct scaling analysis and check the ubiquity of our findings, we first build a 2D microchannel model for transport model benchmarking and scaling against dimensionless groups. Then, synthetic 2D porous media and 3D real sandstone obtained from micro-CT images are also built to validate the existence of transport regimes in a porous media with varying pore sizes. This subsection will introduce how these porous media are built and their basic characteristics.

\subsubsection{Microchannel Model}

\begin{figure}[htp!]
  \centering
    \includegraphics[width=0.7\textwidth]{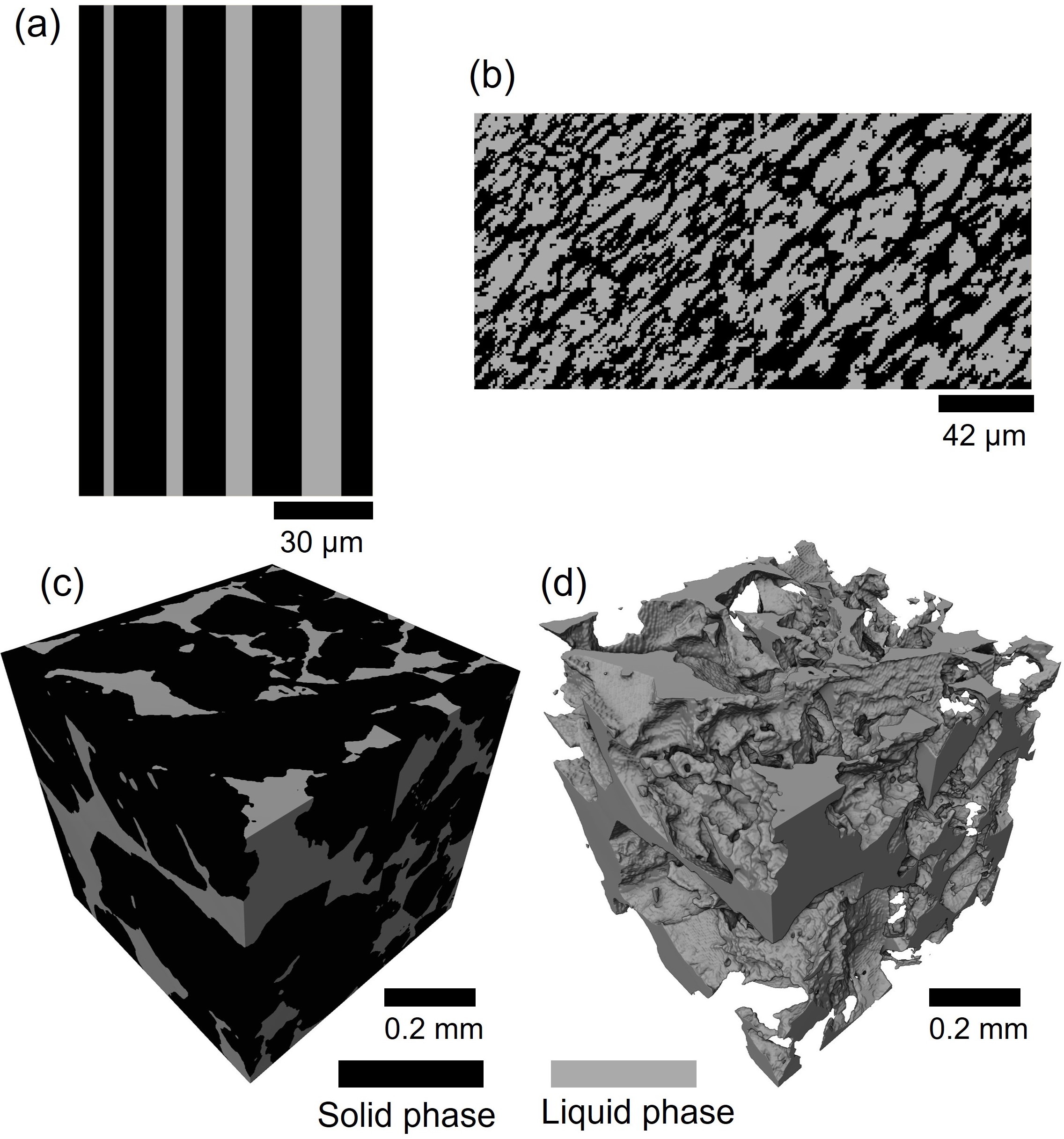}
    \caption{Synthetic porous media (a) 2D microchannel model, resolution = 1$\mu$$m$; (b) 2D synthetic porous media with different pore size, resolution = 1$\mu$$m$; (c) Binary segmentation of 3D concatenated micro-CT image of the bentheimer sandstone, resolution = 2.5$\mu$$m$; (d) Pore space of the bentheimer sandstone.}
    \label{fig:ThreePorousMeidum}
\end{figure}

We start with a homogeneous (the sizes of all flow paths are constant) 2D microchannel model to benchmark the ion transport regime. Herein, a microchannel model with the size of $90\times150$ pixels is created consisting of four microchannels with channel apertures of 3, 4, 7, and 11$\mu$$m$, as shown in Figure \ref{fig:ThreePorousMeidum} (a). The 4 microchannels have the same length. Thus, the channel size is the only factor influencing ion transport.

\subsubsection{Synthetic 2D Porous Model}

The microchannel model provides constant pore size and does not permit the ion flux transfer from one pore to another. To build a porous media with designated pore radius, porosity, and the contrasting between large and small pores in the porous media, we use a sequential indicator simulation algorithm \cite{deutsch1992geostatistical} to build porous media with various pore sizes. Then stick the synthetic media with different pore sizes together to get a heterogeneous (in terms of pore sizes) porous media. During sequential indicator simulation, categorical indicators of 'pore' and 'rock' are populated to each grid (pixel) of the domain based on statistics of pore spaces, including the length of pore space in major and minor directions and the azimuth angle of pore spaces to the vertical spaces. As such, the size of populated pore spaces will be within a specified range and show certain contrasts when generated porous media are combined.

Two porous media with small (3$\mu$$m$) and large (5$\mu$$m$) average pore sizes are generated. We then combine the small-pore media with the large-pore media to get a heterogeneous synthetic porous media model in terms of pore size distribution. Shown as the Figure \ref{fig:ThreePorousMeidum} (b), the 2D model has two sections of the overall size of $256\times127$ pixel with a pixel resolution of 1$\mu$m. The porous media on the left has narrow pore spaces (saturated with liquid phase) connected with each other from top to bottom, ensuring the connectivity between the boundaries and each pore. Another porous section on the right shows wider pore spaces than the left and presents a similar connecting pattern between boundaries and each pore. Such a model will facilitate the observation of ion flow behavior in both small and large pore spaces.   

\subsubsection{3D Porous Sandstone}

We further use a 3D micro-CT image of a Bentheimer sandstone with a voxel resolution of 2.5$\mu$m to explore the ion flow regime in a common type of porous geological rock \cite{kuhn1998determination,de2015kaolinite,mohamaden2016application}. To generate a heterogeneous sandstone image in terms of pore size distribution, we first cut a subdomain with a voxel size of $128\times256\times256$. After that, the corresponding length of the subdomain is increased by two times, resulting in an increase in voxel size from 2.5$\mu$$m$ to 5$\mu$$m$ (the number of voxel increases from $128\times256\times256$ to $256\times512\times512$). A $128\times256\times256$ voxel domain is cropped from the $256\times512\times512$ domain and concatenated to the raw $128\times256\times256$ domain to generate a final domain with $256^{3}$ voxel. Finally, by assuming a constant voxel resolution of 2.5$\mu$$m$ to the entire domain, we generate a 3D heterogeneous porous sandstone in terms of pore size distribution. The binary segmented 3D heterogeneous image is displayed in Figure \ref{fig:ThreePorousMeidum} (c). The pore size distribution can be found in Figure \ref{fig:ThreePorousMeidum} (d), where one side of the domain has substantially larger pore sizes than the other, which is 14$\mu$m and 10.5$\mu$m, respectively.

\subsection{Scaling Analysis of Dimensionless Numbers}

To conduct scaling analysis of the ion flow regime under diffusion, advection, and electromigration, two dimensionless numbers are used to characterise the regimes: (1) $Pe$, defined as the ratio of ion advection flux to net diffusive flux, as defined in Eq.\ref{Pe}; (2) $EDI$, a dimensionless number that defines the ratio of ion electromigration flux to net diffusive flux, as defined in Eq.\ref{Ek}. 

\begin{equation}
    \label{Pe}
    \begin{aligned}
    Pe &=  \frac{J_{zi}^a}{J_{zi}^d}
    \end{aligned}
\end{equation}

\begin{equation}
 \label{Ek}
    \begin{aligned}
    EDI &=  \frac{J_{zi}^e}{J_{zi}^d}
    \end{aligned}
\end{equation}
Where $z$ refers to the main flow direction. Herein, the main driving forces are applied in the z-direction.

For all simulation experiments, various initial and boundary ion concentrations, electric potential, and hydrostatic pressure gradients are applied to the models. Herein, boundary refers to the model's top (inlet) and bottom (outlet) open boundaries, and the surrounding boundaries of 2D and 3D models have no flow. As a result, the ion transport from the inlet to the outlet boundary is defined as positive, while the opposite is defined as negative. It is noteworthy that in this study, $Pe$ and $EDI$ refer to the directional dimensionless number where both positive and negative directions of ion transport are considered \cite{ray2019peclet}. Therefore, $EDI$ and $Pe$ can be positive or negative based on the flux direction. 

For ionic conditions, we adopt the same ion types as used in back diffusion remediation technology \cite{sprocati2022interplay,chowdhury2017electrokinetic}, where a fixed concentration of potassium permanganate solution (KMnO$_{4}$) mixed with hydrochloric acid (HCl) is injected from the inlet boundary, and the domains are initially saturated with sodium chloride solution (NaCl) and HCl. At the outlet boundary, the same concentration as the initial concentration of NaCl mixed with HCl solution is used. The boundary and initial concentration used for all simulations are summarized in Table S1. Other simulation parameters, such as ion diffusivity, are summarised in Table S2 in Supporting Materials. It is noted that all the following results and analyses are based on the permanganate ion that is negatively charged. 

Permanganate ion fluxes into the models under three transport mechanisms correspond to diffusion, electromigration, and advection, respectively. With the initial and boundary condition of ion flow applied herein, the net diffusive flux is always from inlet to outlet. The electromigration flux only depends on the electric potential setting. The permanganate ion moves to the outlet if the anode is set at the outlet and vice versa. The advection flux of ions is caused by the fluid flow, mainly containing two driving mechanisms, one is the fluid flow caused by hydrostatic pressure, and another term is the fluid flow caused by EOF. EOF occurs due to the existence of an EDL when the electric potential is applied to a charged material's surface. The thickness of EDL is characterised by Debye length. In our simulation, when the ionic concentration is set as shown in Table S2, the Debye length is around 4$nm$ which is much smaller compared to the average pore size of all three domains \cite{zhang_electro-osmosis_2017}. Therefore, the EDL is assumed as the thin layer and the HS equation (Eq.\ref{eq:HSequation}) and zeta potential are used to describe the electroosmosis. Local zeta potential at the solid surface is strongly dependent on pH and weakly dependent on ionic concentration \cite{zhang_electro-osmosis_2017}. Considering that our simulations are all performed under the constant pH value (pH=6), the zeta potential value of solid (quartz) is assumed to be constant at -20mV \cite{junior2014behavior}. This assumption is reasonable in situations where the electrical double layer (EDL) is unresolvable and the external transport mechanisms are applied. The driving force on ions due to zeta potential is governed by electroosmosis, which is significantly smaller when compared to electromigration, diffusion, and advection influenced by hydrostatic pressure \cite{tang2023pore}. At each time step, ion fluxes due to the three transport mechanisms and net ion flux are calculated, from which two dimensionless group numbers are calculated, and ion flux flow regimes will be identified by plotting net ion flux on the simulation models.

\section{Results and Discussion}
\label{sec: Results and Discussion}

\subsection{Ion Flow Regime Benchmark}

The ion flow regimes are identified and defined according to visual observations on net ion flux in the synthetic model. In this subsection, we simulate the advection-diffusion-electromigration flow of ions in the 2D microchannel model. We identify four ion flow patterns from the simulation results and benchmark the ion transport regimes accordingly.

\begin{figure}[htp!]
  \centering
    \includegraphics[width=0.8\textwidth]{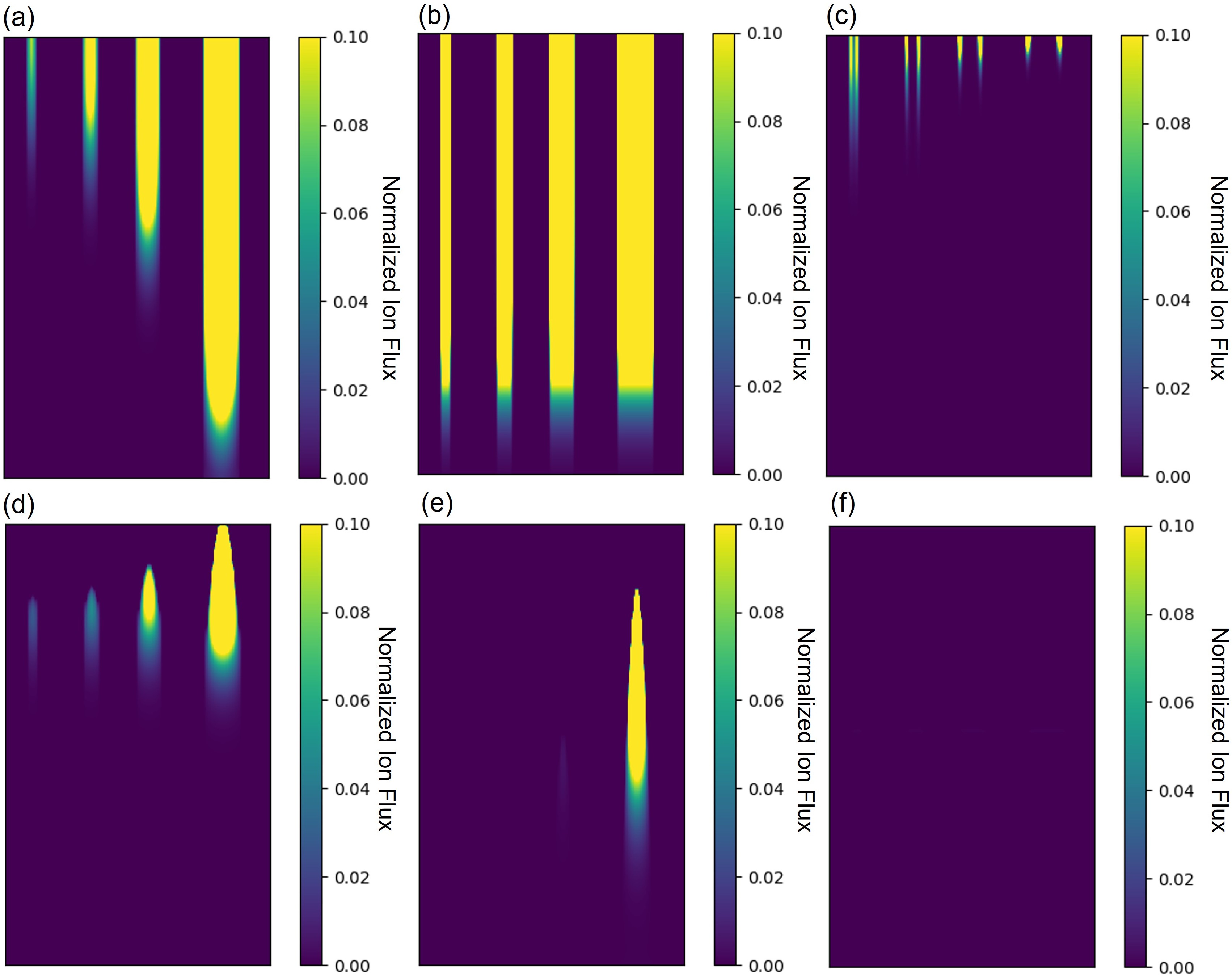}
    \caption{Typical net ion flux transport regimes identified in 2D microchannel model  (a) Large channeling, $EDI = 0, Pe = 29.61$; (b) Uniform flow, $EDI = 1.93, Pe = -9e-4$; (c) Small channeling, $EDI = 9.5e-6, Pe = -0.95$; (d) Large channeling, $EDI = -2.5, Pe = 1$; (e) No flow, $EDI = -20.78, Pe = 10.84$; (f) No Flow, $EDI = -2, Pe = 1e-3$.}
    \label{fig:FourFlowRigimes}
\end{figure}

Large channeling occurs when ion flux is advection dominant. This occurs when the $Pe$ is positive and relatively large compared to $EDI$, meaning that the hydraulic pressure is applied from top to bottom of the model and is the major transport mechanism compared to diffusion and electromigration. With large channeling, ions prefer transporting through the larger channels with less transport or bypassing channels. The typical transport pattern is shown in Figure \ref{fig:FourFlowRigimes} (a).

Additionally, when an electric potential is applied with the anode at the top and cathode at the bottom of the 2D microchannel model, meaning $EDI$ is negative, the permanganate ion has a uniform electromigration flux towards the anode (from bottom to the top). Under the condition where advection predominates over electromigration, this uniform electromigration flux will have less influence on the ion transport in large channels, but the advection flux in small channels may be canceled out by the uniform electromigration flux. The overall ion transport pattern becomes more heterogeneous as a result. Moreover, such a flow pattern will start to change again when the electric potential gets strong enough compared to the advection. In this condition, there is no net ion flux through the small channels, and the net ion flux in the large channel reduces but still exists, shown as figure \ref{fig:FourFlowRigimes} (d). We define all these transport patterns as large channeling.

A uniform flow pattern occurs where $EDI$ dominants over $Pe$ or both are too small compared to the net diffusive flux. A typical uniform flow pattern is presented in Figure \ref{fig:FourFlowRigimes} (b), uniform flow delivers high sweep efficiency compared to large channeling (both small and large channels are swept by ion fluxes). For an advection-diffusion system, uniform flow only occurs when diffusion dominates the transport. With the external electric potential, uniform flow can occur with both electromigration and diffusion. One can alter the ion concentration and electric potential to change the flux of the uniform flow. The channel length or tortuosity between large channels and small channels is also a factor that affects the ion transport regime; however, it is ignored in the 2D microchannel model, where all channels have the same length. The impact of tortuosity is discussed in the synthetic 2D porous media and 3D sandstone domain.

Small channeling prevails as negative $Pe$ increases and becomes the main transport mechanism over $EDI$. Applying a large pressure gradient on the 2D microchannel in the opposite direction (from bottom to the top) to the uniform ion fluxes will counterbalance the ion fluxes in the large channels first. Consequently, since the negative flux by pressure is larger in the middle of the channel and smaller along the solid surface due to viscous force, the negative net ion flux starts to occur at the middle part of the large channel. The typical small channeling pattern is shown in Figure \ref{fig:FourFlowRigimes} (c). When the $EDI$ becomes negative, advection and electromigration flux counterbalance the net diffusive flux. Small channeling can still exist when the net diffusive flux is comparable to negative advection flux and electromigration flux. However, no flow appears to occur with further increase of the advection and electromigration flux. In some cases of large and small channeling flow regimes, ion fluxes in the minor transport paths (small channels during large channeling or large channels during small channeling) would first disappear when they meet counterbalance flow mechanisms in the opposite direction. A transition flow pattern is defined between small channeling the no flow pattern. Ion fluxes in all channels (flow paths) will disappear once the counterbalances dominate. Herein, we define both zero net ion flux and negative net ion flux as the no flow regime. Figure \ref{fig:FourFlowRigimes} (e) and (f) show no flow regime originating from large channeling and uniform flow, respectively. As such, we define these flow behavior observations as no flow.

\begin{figure}[htp!]
  \centering
    \includegraphics[width=0.8\textwidth]{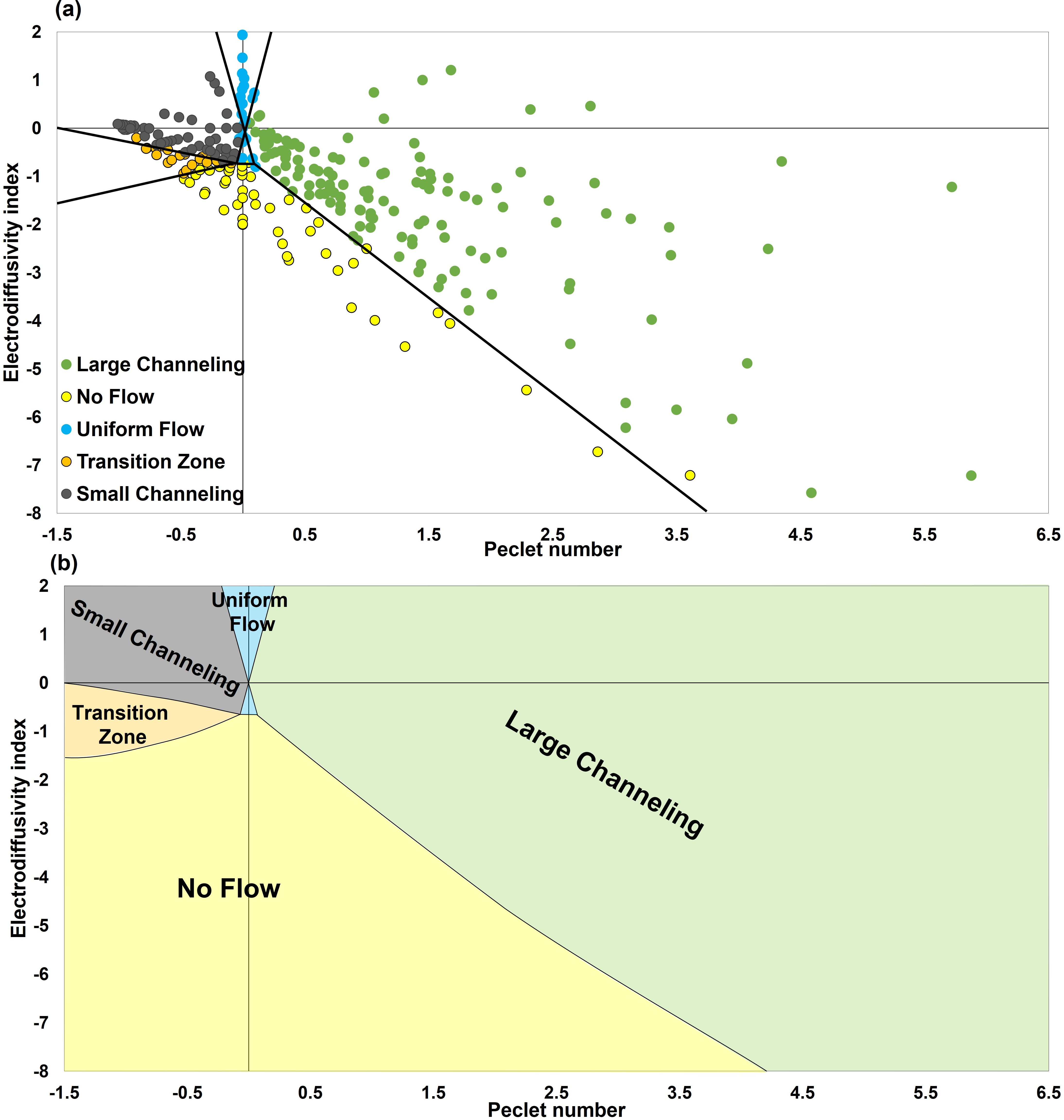}
    \caption{Illustration of advection-diffusion-electromigration flow regime. (a) Ion transport pattern with different P\'eclet number and Electrodiffusivity index in the 2D microchannel model. (b) Ion transport regime defined from (a).}
    \label{fig:BenchMarkFlowRegime}
\end{figure}

After understanding the ion flow patterns in the advection-diffusion-electromigration system, we run 40 simulations using various combinations of $Pe$ and $EDI$ on the 2D microchannel model and defining the transport pattern based on visual observation. The simulation results are represented graphically in Figure \ref{fig:BenchMarkFlowRegime} (a). Based on the simulation results, the ion transport regime is defined in Figure \ref{fig:BenchMarkFlowRegime} (b). A positive value of the dimensionless number means the driving force moves ion from inlet to outlet. 

In the first quadrant, $EDI$ and $Pe$ are both positive. When $Pe$ is large enough, the transport would always be large channeling, while when $EDI$ is dominant and $Pe$ is small, a uniform flow can be achieved. In the second quadrant, $EDI$ is positive and $Pe$ is negative, and a symmetrical uniform flow region exists where $EDI$ is dominant and $Pe$ is small. Small channeling is located where negative $Pe$ refers to the counterbalance of flux in large channels. In the third quadrant, the $EDI$ and $Pe$ are both negative; when the net diffusive flux can still counterbalance the negative flux caused by negative $EDI$ and $Pe$, small channeling still exists. With the further increase of the $Pe$, the flux region in the large channel has further reduced, which increases the diffusion to the region that still has flux (the flow along the solid surface), such as the small channeling in Figure \ref{fig:FourFlowRigimes} (c), where the ion flux along the solid surface of the larger channel is larger than that in a small channel. At a certain $Pe$ value, the flow in the small channel stopped, while there is still a flux along the solid surface in the large channel. We name this region the transition zone. With the further increase of the $EDI$ and $Pe$, no flow pattern starts to appear. Finally, in the fourth quadrant, the $EDI$ is negative and $Pe$ is positive. No flow pattern happens when the electromigration is more dominant than the advection and diffusion. When advection starts to dominate, large channeling occurs again. A uniform flow region in the third and fourth quadrants is defined when the net diffusive flux and electromigration flux are counterbalanced; however, the net diffusive flux is still larger than the electromigration flux.

\subsection{Observation in Heterogeneous Domains}

After benchmarking the ion transport regime on a 2D microchannel model, more complex porous media with different tortuosity are used to investigate the ion flow in heterogeneous domains under advection-diffusion-electromigration. 

\subsubsection{Ion Flow Regime on Synthetic 2D Porous media}

\begin{figure}[htp!]
  \centering
    \includegraphics[width=1\textwidth]{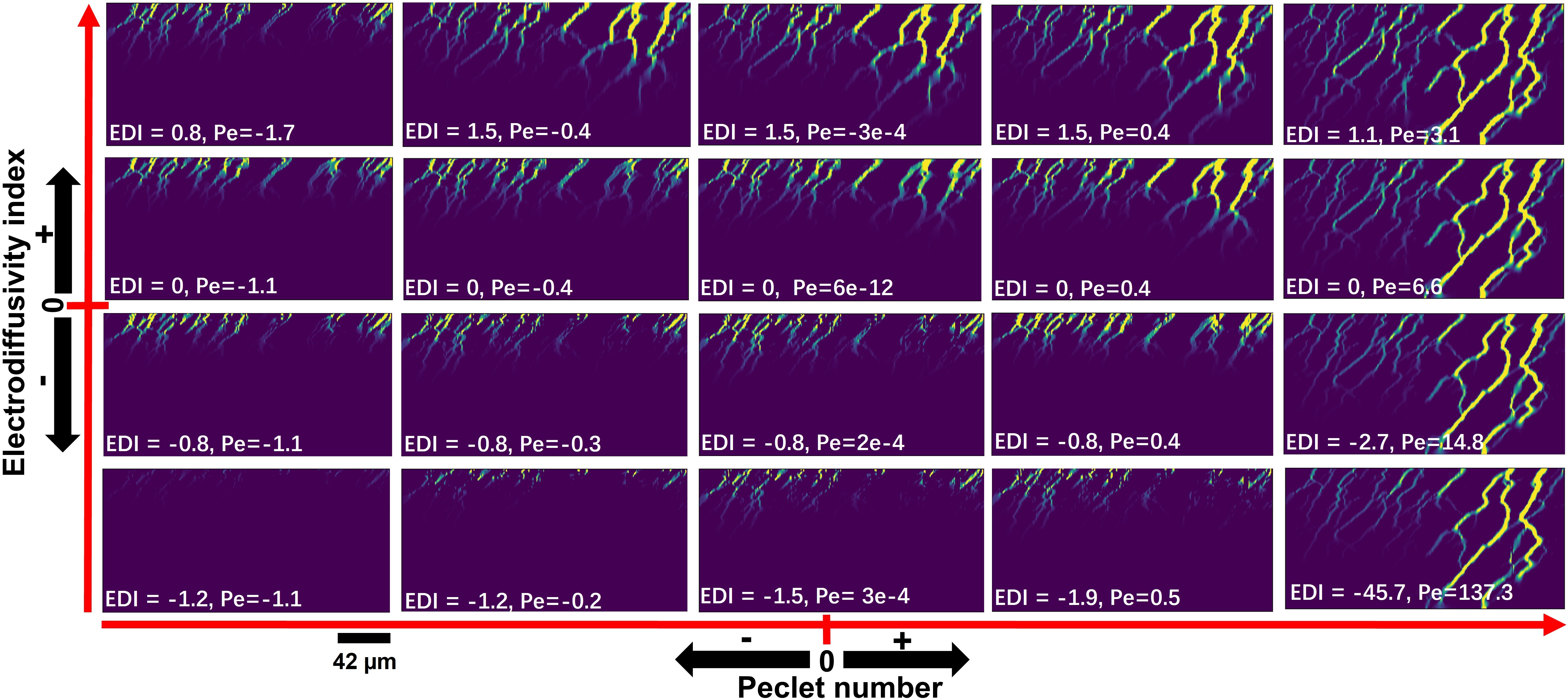}
    \caption{Ion transport regime for the 2D heterogeneous porous medium. All ion transports presented here are at time = 0.51sec. The ion flux in each subplot is normalized to 0-1 within itself.}
    \label{fig:2DFlowRegime}
\end{figure}

To generate the ion flow regime for the 2D Porous media, we perform 20 simulations that contain $EDI$ and $Pe$ combinations in four quadrants. Figure \ref{fig:2DFlowRegime} shows the transport regime of 20 simulations at the same timestep. We observe four flow regimes: large channeling, uniform flow, small channeling, and no flow. Large channeling can be observed at the right column where positive $Pe$ occurs. With decreasing $EDI$ from positive to negative (from 1.1 to -45.7), the flow in the low porosity zone is counterbalanced by electric potential, making the large channeling more apparent. These observations of the large channeling regime fit with the benchmark transport regime defined in Figure \ref{fig:BenchMarkFlowRegime} (b).  By decreasing the $Pe$ to around 0.4, large channeling becomes less apparent. Meanwhile, with decreasing $EDI$, more uniform flow can be observed, such as the one with $EDI = -0.8$ and $Pe = 0.4$. When the $Pe$ reduces further, the ion transport is dominated by diffusion or electromigration and is counterbalanced by pressure in the negative $Pe$ quadrant. Therefore, the uniform flow and small channeling would occur ($EDI = 0$, $Pe = 6e-12$ and $EDI = -1.5$, $Pe = 3e-4$). For the negative $Pe$ and $EDI$ regions, the transport regimes are mainly small channeling and no flow ($EDI = -0.8$, $Pe = -1.1$ and $EDI = -1.2$, $Pe = -1.1$), which matches the fourth quadrant of our benchmark transport regime. Additionally, with negative $Pe$ and large positive $EDI$, small channeling is also presented ($Pe = -1.7$ and $EDI = 0.8$), which also matches the third quadrant transport regime in Figure \ref{fig:BenchMarkFlowRegime} (b). It is noted that the boundary of the transport regime in the synthetic 2D porous media is different from that obtained in the 2D microchannel due to the influence of the tortuosity variation in the synthetic 2D porous media. A good example is when $EDI = 0$ and $Pe = 6e-12$, a uniform flow is observed with a 2D microchannel where tortuosity remains the same for each channel; meanwhile, a weak large channeling is achieved with the synthetic 2D porous media as shown in Figure \ref{fig:2DFlowRegime}.
 
\begin{figure}[htp!]
  \centering
    \includegraphics[width=1\textwidth]{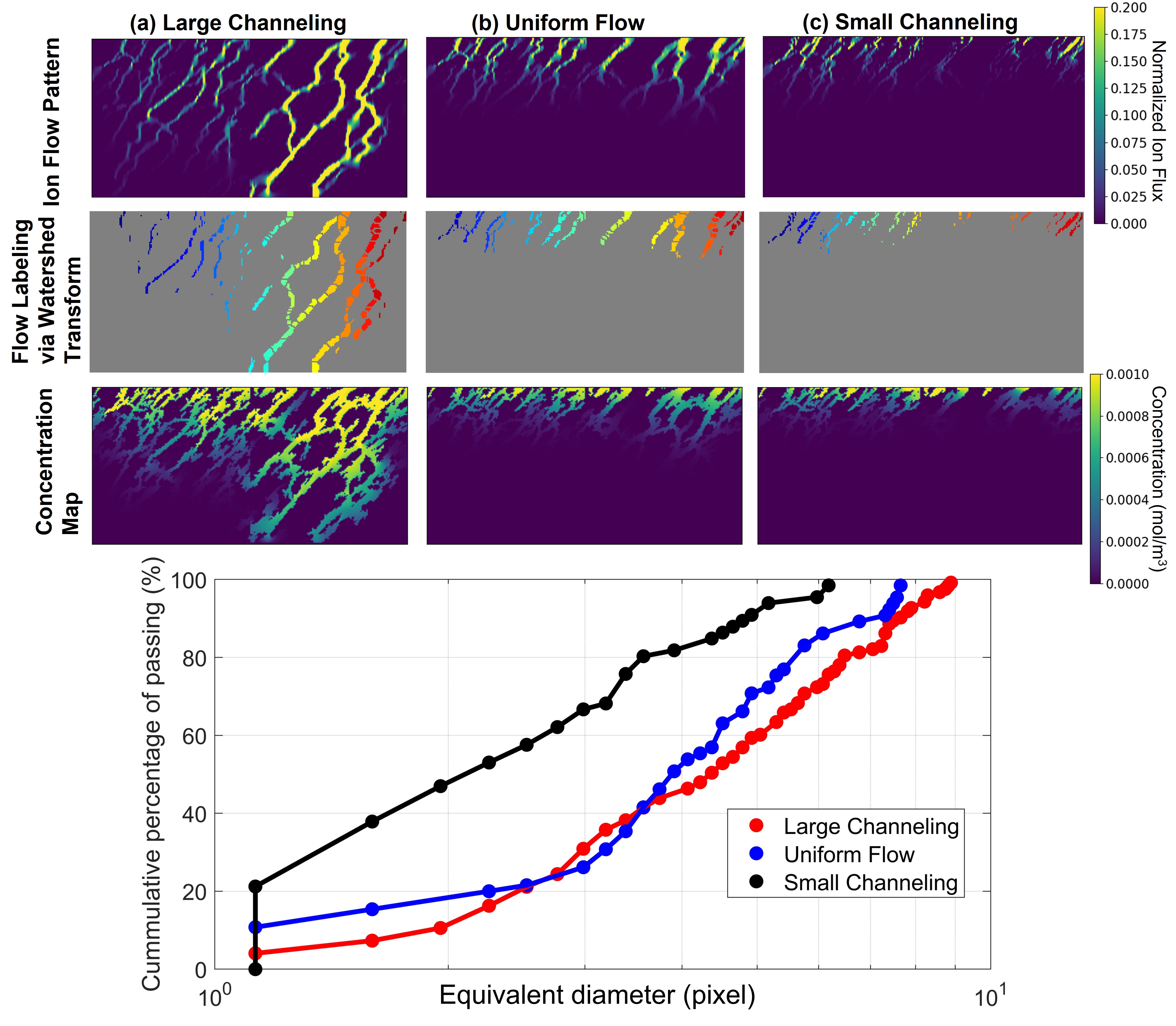}
    \caption{Occupied pore size distribution and the visualization of concentration map of three typical transport regimes in the advection-diffusion-electromigration system, (a) $EDI = 1.1, Pe = 3.1$; (b) $EDI = 0, Pe = 6e-12$; (c) $EDI = 0.8, Pe = 1.7$. The pore size is generated using the watershed transform algorithm. The flowed pore size is calculated using the equivalent diameter method and plotted in the unit of pixel.}
    \label{fig:FlowSizeDis}
\end{figure}

To further compare and quantify these transport regimes, we analyze the occupied pore size distribution and the concentration map for a typical large channeling, uniform flow, and small channeling at the same timestep, as shown in Figure \ref{fig:FlowSizeDis}. Here, we study the size of pores that are occupied by ions with a concentration greater than 1\% of the inlet concentration\cite{jangda2023pore}. Based on the ion flux pattern, we use the watershed transform algorithm to separate each pore that is occupied by ions \cite{roerdink2000watershed}. We then calculate the pore size based on the equivalent diameter. From the size distribution map, we find that greater than 64\% of the ion-occupied pores for small channeling have the equivalent diameter of less than 3 pixels, which is twice that for uniform flow and large channeling. The maximum flow diameter for small channeling is around 6 pixels. However, 16\% ion-occupied pores for uniform flow and 28\% ion-occupied pores for large channeling have diameters greater than 6 pixels. For uniform flow, the maximum flowed pore size is around 7.8 pixels. At 3.8 pixels in size, which is half the maximum size, 50\% of the pores are greater than that, and the other 50\% are smaller. Therefore, the pore size distribution map of each flow regime can be used to quantify the transport regime we visually defined.

\begin{figure}[H]
  \centering
    \includegraphics[width=1\textwidth]{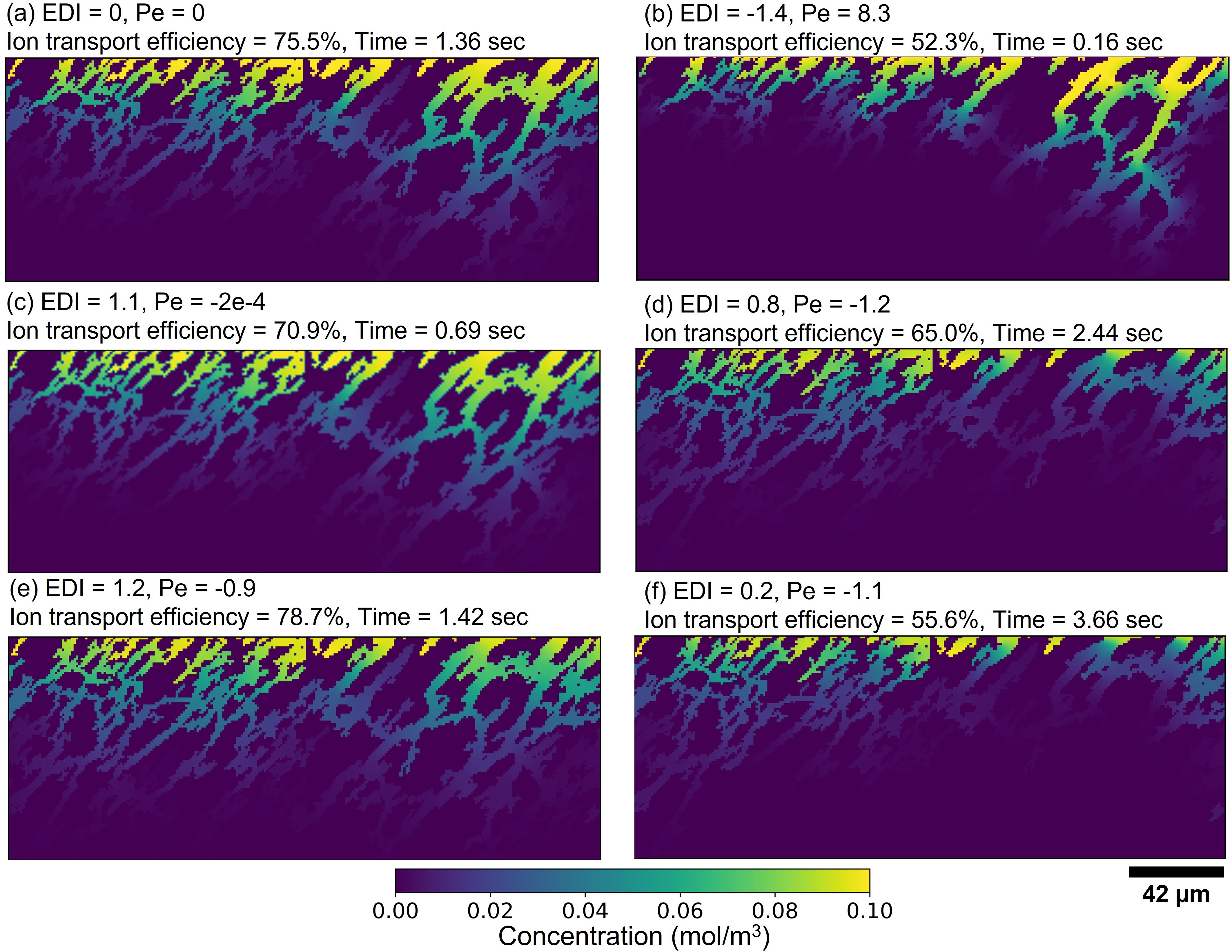}
    \caption{Ion transport regimes and the ion transport efficiency at breakthrough times. Large channeling, uniform flow, and small channeling are observed with different $EDI$ and $Pe$ combinations.}
    \label{fig:BTFlowSizeDis}
\end{figure}

In addition, we studied the transport regimes up until the breakthrough. Which is defined by when 1\% of the inlet ion concentration is observed at the outlet, it is defined as breakthrough. 6 different cases including small channeling, uniform flow, and large channeling are simulated until the breakthrough. The results are reported in Figure \ref{fig:BTFlowSizeDis}. When both $EDI$ and $Pe$ are zero, the ion transport is diffusion only, and uniform flow should occur. However, due to the tortuosity of the porous media, the flow path for ions in the larger pore region is shorter than the smaller pore regions; thus, the breakthrough occurs at the outlet of the large pore region \cite{shen_critical_2007}. The ion transport efficiency as a quantification parameter is the ratio of the pore volume with an ion concentration larger than 1\% of the inlet ion concentration to the overall pore volume. The ion transport efficiency is 75.5\% at time = 1.36sec for diffusion flow. By giving a large positive $EDI$ compared to negative $Pe$, such as the case in Figure \ref{fig:BTFlowSizeDis} (c), the movement of ion increases, and the breakthrough time decreases to 0.69sec. Due to the early breakthrough, the ion transport efficiency also decreases. By increasing the negative $Pe$ to be comparable to $EDI$, as shown in case Figure \ref{fig:BTFlowSizeDis} (e), the fast movement of ion in the large pore region by tortuosity is counterbalanced by negative advection flow. Therefore, a more uniform flow is formed with the highest ion transport efficiency (83.2\%) as well as a longer breakthrough time (1.83sec) is observed. In another case where advection is dominant and $EDI$ is negative (Figure \ref{fig:BTFlowSizeDis} (b)), an apparent large channeling is formed, and the breakthrough occurs at t = 0.16sec at the outlet of large pore region with the lowest ion transport efficiency (52.3\%). Furthermore, small channeling can be observed when $EDI$ is slightly less than $Pe$, such as in Figure \ref{fig:BTFlowSizeDis} (d), the negative advection flux counterbalanced the ion flox in larger pore region, and thus the flow breakthroughs over the smaller pore region. The ion transport efficiency is 65.0\%, but more time is required for the ion to move till the breakthrough (t=2.44sec) compared to large channeling cases. With the further decrease of $EDI$, the negative advection flux becomes dominant compared to positive diffusion and electromigration (Figure \ref{fig:BTFlowSizeDis} (f)), apparent small channeling is formed, and ion transport efficiency decreases. The ion moves slower with less $EDI$ and thus the time is longer to reach a breakthrough. Based on these cases, we find that by using the electric potential as a component of the driving force in the advection-diffusion system, any of the desired ion transport behavior can be achieved, which provides flexibility when dealing with different applications.

\subsubsection{Ion Transport Regime on a 3D Sandstone}

\begin{figure}[htp!]
  \centering
    \includegraphics[width=0.9\textwidth]{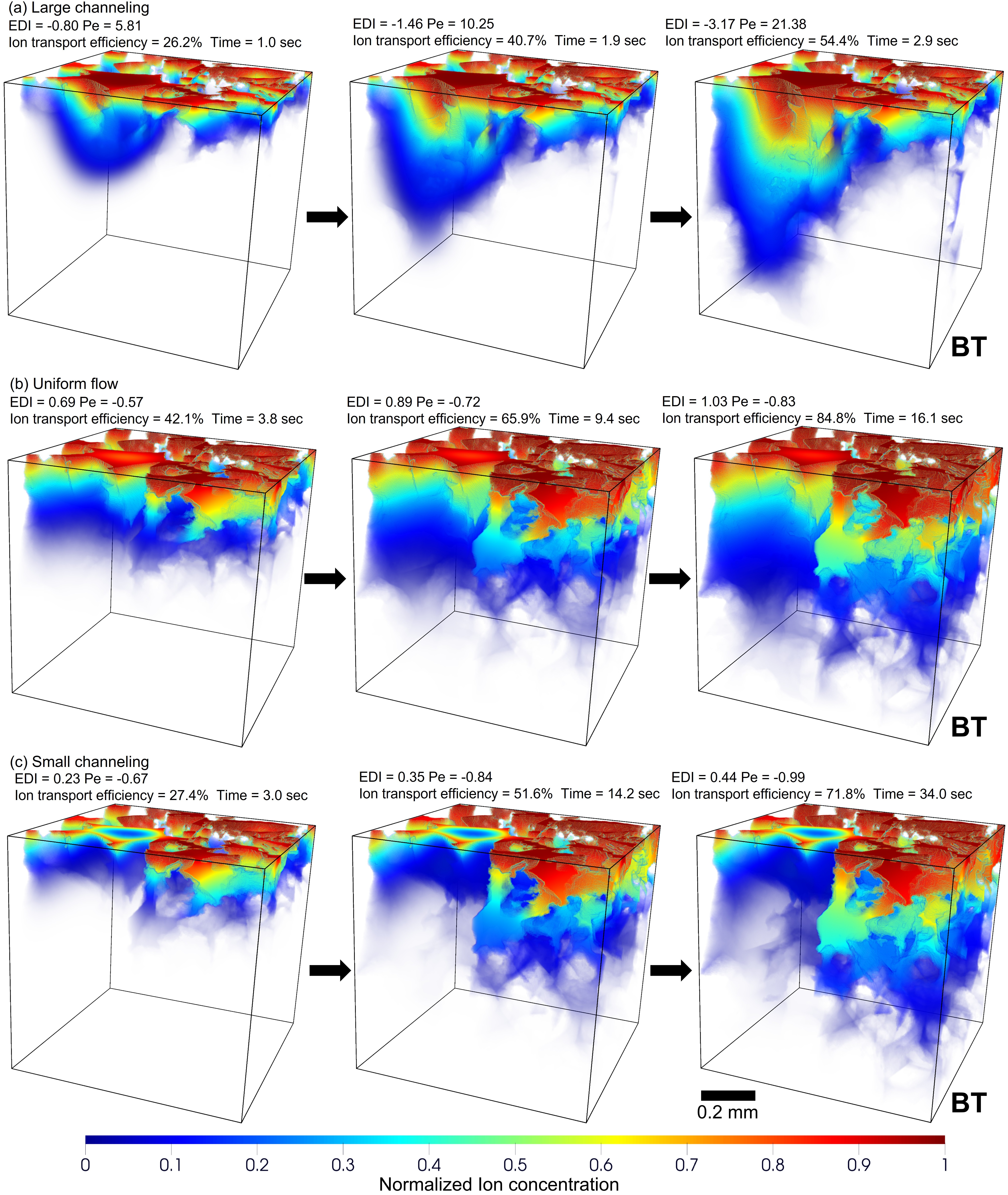}
    \caption{Ion transport regime for the 3D sandstone micro-CT image (a) Large channeling; (b) Uniform flow; (c) Small channeling. For each transport regime, three timesteps are displayed, and the simulation stops when 1\% of the ion concentration reaches the outlet (last column), referring to a breakthrough (BT). The ion concentration in each subplot is normalized to 0-1.}
    \label{fig:3Dsimulation}
\end{figure}

The ion transport regime is also tested on a 3D sandstone micro-CT domain. For the 3D micro-CT domain, the porosity and mean equivalent diameter for the subdomain with larger pores is 0.36 and 14$\mu$$m$, while that for the subdomain with smaller pores is 0.27 and 10.5$\mu$$m$. Three cases including large channeling, uniform flow, and small channeling are simulated until 1\% of inlet ion concentration reaches the outlet (breakthrough). The results are shown in Figure \ref{fig:3Dsimulation}. Large channeling occurs in Figure \ref{fig:3Dsimulation} (a), where $EDI$ is negative and $Pe$ is positive and dominant. The majority of ions move through the high porosity region and reach the breakthrough after 2.9sec with the final ion transport efficiency of 54.4\%. The paired $EDI$ and $Pe$ of three timesteps fall into the large channeling flow regime in the fourth quadrant, as benchmarked in Figure \ref{fig:BenchMarkFlowRegime}. For Figure \ref{fig:3Dsimulation} (b), the direction of applied hydrostatic pressure and electric potential are reversed compared to Figure \ref{fig:3Dsimulation} (a), resulting in a positive $EDI$ and negative $Pe$. A uniform flow pattern is observed where ions are uniformly distributed. The breakthrough time is delayed compared to large channeling because the ions move slower with negative $Pe$. The ion transport efficiency at breakthrough for uniform flow is 84.8\%. In the last case (Figure \ref{fig:3Dsimulation} (c)), the hydrostatic pressure increases and electric potential decreases compared to Figure \ref{fig:3Dsimulation} (b), resulting in small channeling. By comparing the small channeling case to the large channeling case, we discover that ions do not move through the main flow channel in Figure \ref{fig:3Dsimulation} (a) and instead move through the subdomain with low porosity. For ion breakthrough, small channeling takes more time than the other two flow patterns. From the case of small channeling flow, the simulation results fall within the third quadrant of the ion transport regime presented in Figure \ref{fig:BenchMarkFlowRegime}  Overall, from the simulation on the 3D sandstone micro-CT domain, we also confirm that the observed ion transport regime matches the benchmarked transport regime in Figure \ref{fig:BenchMarkFlowRegime}.

\subsection{Implication on Engineering Problems}

Electrokinetic transport can be applied to a subsurface system to improve the efficiency of ion transport. We find that ion sweep efficiency is high when electrokinetic transport is dominant, e.g. uniform flow, see \ref{fig:BTFlowSizeDis} (c,d,e) and \ref{fig:3Dsimulation} (b,c); or small channeling, see \ref{fig:BTFlowSizeDis} (f) and \ref{fig:3Dsimulation} (c). Such sweep improvement is achieved through the controllable preferential transport path of ion harnessing the interaction between the Electrodiffusivity index and P\'eclet number. In field implementations, the magnitude of the applied electric potential will be of great importance \cite{sprocati2019modeling}. According to our transport regime map, one can easily find the appropriate range of $EDI$ based on hydrostatic pressure, pore size variation, and desired transport regime, no matter whether the target zone is located in a low permeable zone (small channeling), high permeable zone (large channeling), or both (uniform flow). Thus, our study theoretically proves the feasibility of controlling the path and direction of ion transport. Achieving such control requires both hydrostatic pressure and electric potential to be applied. In certain cases, these fields may be applied in opposite directions to each other to control transport patterns. This implies the active use of work to impede the overall solute flow in a porous system in exchange for a more stable flow front through heterogeneity. The engineering applications of such control are wide-reaching. 

In our transport regime map, the small channeling and uniform flow, in most cases, involves advection in the opposite direction of electromigration to counterbalance the ion fluxes in large channels of the porous media. This is intuitively a less efficient system than its counterpart since energy is intentionally wasted. However, in practice, stringent ion flux control is in demand, e.g. for water management and soil remediation. In underground water remediation, the specific location and dimension of the pollution source are usually unknown \cite{mackay_groundwater_1989,chowdhury2017electrokinetic,lima2017environmental}. Hence, a complete cut-off connection between pollution sources and water production wells is desired when returning the contaminated aquifer to drinking-water quality. In this condition, stringent control of ion flows in the subsurface is more important than the technical efficacy of the implementation scenario \cite{mackay_groundwater_1989}. Therefore, the electrokinetic-advection-diffusion transport system proposed in this study and the interplay between electromigration and advection during small channeling and uniform flow regimes may pose advantages in these circumstances. 

Besides small channeling and uniform flow, large channeling also needs careful design. For mineral resource recovery technologies, e.g. for in-situ recovery, perfect ion flow control is no longer in demand. Instead, the sweep and time efficiency are of greater importance \cite{martens_toward_2021,wang2022green}. At the initial stage of recovery, large channeling with hydrostatic pressure-induced advection is desired to efficiently produce the resources which are distributed along and beneath the large channels' surface. Afterward, maintaining high hydrostatic pressure will no longer bring a better sweep of ion flows. Adjusting the electric potential based on the flow regime map to improve the sweep and transport of ions will bring further economic benefits to the project. 

Overall, the observed transport regime map results from the complicated interplay between electromigration, advection, and diffusion, which is readily applicable to various engineering problems. The relation between the three mechanisms may vary over time and space (heterogeneity) in the process of implementation. One needs to combine our proposed flow regime map and the real condition to design the expected ion transport scenarios (could be one or a set of flow regimes over each stage of the engineering project). Here we only consider three transport mechanisms and porous media with contrasting pore sizes; future work on the impact of chemical reactions on the corresponding flow regimes at various scales (e.g. meter scale) will complement this scaling analysis on ion transports in an advection-diffusion-electromigration transport system.

\section{Conclusions}
\label{sec:conclusions}

We performed a fundamental pore-scale model-based scaling analysis on ion transport under an advection-diffusion-electromigration system in three image-based domains, including a homogeneous (constant aperture along flow paths) microchannel domain, and heterogeneous (contrasting pore sizes over simulation domain) 2D and 3D domains. The simulations are performed using our open-source solver OPM/LBPM \cite{mcclure2021lbpm}. Under advection-diffusion-electromigration conditions,  ion transport regimes can be benchmarked using the P\'eclet number, $Pe$, and a newly defined dimensionless number, $EDI$. Four ion transport regimes are defined: large channeling, uniform flow, small channeling, and no flow. We propose that the ion transport can be well-controlled by changing the applied electric potential, and hydrostatic pressure, and injecting ion concentration. A generated 2D domain and a 3D sandstone micro-CT image are then used to demonstrate the observed ion transport regimes in heterogeneous porous media. The simulation results highlight that the defined transport regimes exist in heterogeneous porous media, but the boundary between each transport regime may vary depending on heterogeneity. Based on these findings, the advection-diffusion-electromigration transport system has the potential for many engineering problems, including, e.g., underground water management \cite{malaeb2011reverse,liu2017effects} and underground resource recovery \cite{karami2021review,tang2023pore}, where the pollutants or target zone can be uniformly distributed in a geological formation or mainly located in the low permeable region.

\section{Acknowledgment}
\label{sec:Acknowledgement}
The authors acknowledge the Tyree X-ray CT Facility at the Mark Wainwright Analytical Centre (MWAC) of UNSW funded by the UNSW Research Infrastructure Scheme, for the acquisition of the 3D micro-CT images.

\bibliographystyle{ieeetr}  
\bibliography{LBPM_COMSOL_paper} 

\end{document}


\setcounter{tocdepth}{4}
\setcounter{secnumdepth}{4}
\maketitle

\pagebreak

\renewcommand\arraystretch{1.5} 
\begin{table}[htp!]
\setlength{\tabcolsep}{0.8mm}{ 
\centering
\caption{Summary of the simulation ionic concentration at inlet, outlet, and initial condition for all three domains. pH, Porosity, domain resolution, and zeta potential are also listed. }
\begin{tabular}{l|lll|lll|lll|l|l|l}
\hline
                     & \multicolumn{2}{c}{\textbf{Initial electrolyte}} &    & \multicolumn{2}{c}{\textbf{Injection}} &    & \multicolumn{2}{c}{\textbf{Outlet}} &    & \textbf{Porosity} & \textbf{\begin{tabular}[c]{@{}l@{}}Resolution \\($\mu$m)\end{tabular} } & \textbf{\begin{tabular}[c]{@{}l@{}}Zeta potential \\ (mV)\end{tabular}} \\
Domain               & \begin{tabular}[c]{@{}l@{}}NaCl \\ (mM)\end{tabular}   & \begin{tabular}[c]{@{}l@{}}HCl \\ (mM)\end{tabular}          & pH & \begin{tabular}[c]{@{}l@{}}KMnO$_{4}$ \\ (mM)\end{tabular}     & \begin{tabular}[c]{@{}l@{}}HCl \\ (mM)\end{tabular}     & pH & \begin{tabular}[c]{@{}l@{}}NaCl \\ (mM)\end{tabular}    & \begin{tabular}[c]{@{}l@{}}HCl \\ (mM)\end{tabular}    & pH &          &                        &                     \\ \hline
2D Microchannel      & 3                   & 9.9E-04           & 6  & 3              & 9.9E-04      & 6  & 3            & 9.9E-04     & 6  & 0.31     & 1                      & -20                 \\\hline
2D Porous Media      & 3                   & 9.9E-04           & 6  & 1E-03          & 9.9E-04      & 6  & 3            & 9.9E-04     & 6  & 0.49     & 1                      & -20                 \\\hline
3D Sandstone  & 3                   & 9.9E-04           & 6  & 6              & 9.9E-04      & 6  & 3            & 9.9E-04     & 6  & 0.32     & 2.5                   & -20  \\ \hline              
\end{tabular}
\label{tab:concentration}}
\end{table}

\renewcommand{\arraystretch}{1.6}
\begin{table}[htp!]
\centering
\caption{Parameters used in the simulation.}
\begin{tabular}{lll}
\hline
\textbf{Physical parameter}             & \textbf{Symbol}         & \textbf{value} \\ \hline
Density of electrolyte solution         &  $\rho$                       & 998.2 $kg/m^3$    \\
Temperature                             &    $T$                     & 293.15 $K$       \\
Kinematic viscosity of electrolyte solution  &  $\upsilon$                & $1.003\times10^{-6} Pa s$  \\
Fluid dielectric constant               &    $\epsilon_{R}$               & 78.5           \\
Permittivity of vacuum                  &    $\epsilon_{0}$          & $8.85\times10^{-12} F/m$               \\
Diffusivity of H$^+$                    &   $D_{H^+}$                      & $8.59\times10^{-9} m^2/s$      \\
Diffusivity of OH$^+$                    &  $D_{OH^+}$                        & $4.86\times10^{-9} m^2/s$      \\
Diffusivity of Na$^+$                    &  $D_{Na^+}$                        & $1.23\times10^{-9} m^2/s$      \\
Diffusivity of K$^+$                       &  $D_{K^+}$                        & $1.81\times10^{-9} m^2/s$      \\
Diffusivity of Cl$^+$                       &  $D_{Cl^+}$                        & $1.87\times10^{-9} m^2/s$      \\
Diffusivity of MnO$_{4}^{-}$                &  $D_{MnO_{4}^{-}}$          & $1.50\times10^{-9} m^2/s$      \\ \hline
\end{tabular}
\label{tab:parameter}
\end{table}